\documentclass[twocolumn,showpacs,preprintnumbers,prC]{revtex4}
\usepackage{graphicx}
\def\bea {\begin{eqnarray}}
\def\eea {\end{eqnarray}}

\def\be {\begin{equation}}
\def\ee {\end{equation}}
\def\nn {\nonumber}
\begin{document}
\title{Gluon bremsstrahlung by heavy quarks - its effects on transport coefficients and equilibrium distribution}
\author{Surasree Mazumder, Trambak Bhattacharyya and Jan-e Alam}
\medskip
\affiliation{Theoretical Physics Division, 
Variable Energy Cyclotron Centre, 1/AF, Bidhan Nagar ,
Kolkata - 700064}
\date{\today}
\begin{abstract}
The effects of gluon radiation by charm quarks  on the transport 
coefficients {\it e.g.} drag, longitudinal and transverse diffusion 
and shear viscosity have been studied within the ambit of perturbative
quantum chromodynamics (pQCD) and kinetic theory. We found that while 
the soft gluon radiation has substantial effects on the transport coefficients
of the charm quarks in the quark gluon plasma  its effects on the equilibrium 
distribution function is insignificant. 
\end{abstract}

\maketitle

\section{Introduction}
Recently the study of Quark Gluon Plasma(QGP), expected to be created in heavy-ion collisions at the 
Relativistic Heavy Ion Collider (RHIC) and Large Hadron Collider (LHC) energies, intrigues the 
scientific community with its multifarious interesting aspects. 
Therefore, in order to understand different properties of QGP we need to probe it.
Among many, one of the efficient probes is the charm quark (CQ) produced in
the early hard collisions of the partons from the colliding nuclei. 
Generally the transport coefficients 
are sensitive to the interaction of the probes with the medium. 
Hence, estimation of the various transport coefficients of QGP by using CQ is a 
field of high contemporary interest. 
Moreover, strong coupling of the probes with the medium may bring the probe in equilibrium
with the bulk matter, so that the probe follow a momentum distribution similar to that of the
constituents of the medium. In the present work, we will consider the QGP as the
thermal medium of light quarks, their antiparticles and gluons 
and the CQs as probes. This will enable us to estimate the drag,
diffusion and shear viscous coefficients of QGP and the nature of the equilibrium
distribution of the CQs. 
The reasons behind choosing CQ as a probe are:
(i) being created from the early hard collisions, CQ can experience the 
hot/dense medium from its birth. The CQ distribution
function is different from that of the medium particle and
being heavier than the constituent particles of QGP, it 
does not get equilibrated quickly and hence it qualifies to act as the Brownian particle.
(ii) The probability of the production of CQ (with mass M) inside the thermal medium
of temperature $T$ ($T<<M$) 
is very less, hence,  the probability of annihilation of CQs in the 
QGP is also small. Therefore, the CQs witness the entire evolution of the bath.
The probability of creation and annihilation of bottom quarks is even smaller,
therefore, the present work can be extended to the bottom quarks also. 
%Although calculations are performed for charm quark here,
%it can be extended  to case of bottom quark, too. 
While propagating through the QGP the CQ  interacts with the
medium particle via two dominant processes: a)collisional 
or elastic interaction  and b) inelastic interaction like  
gluon bremsstrahlung. In earlier works, while calculating the momentum diffusion
coefficients, either the gluon radiation by the CQ 
have been ignored~\cite{BS,hees,gossiaux,lang} or it is
calculated for non-relativistic CQ~\cite{moore,mooreprl} or estimated 
by first determining the drag from radiative energy loss~\cite{surasree,santosh,abirplb} 
and then using the Einstein relation between drag and diffusion coefficients\cite{vitev}.

In this work, we calculate, using pQCD, the transverse and longitudinal 
diffusion coefficients of the CQ undergoing radiative 
loss by emitting gluons while propelling through the
QGP. %The momentum space diffusion physically means mixing of the high momentum  with
%the low momentum zone to drive the whole system toward a equilibrium state.   
We consider that the emitted gluons, being soft, get absorbed in the medium,
thereby resulting in transporting the energy from the fast moving CQ to the slow moving 
constituents of the bath. This transportation of momentum is reflected in the momentum
diffusion coefficients of the CQ in QGP.

%The issue of HQ drag in QGP has been studied earlier~\cite{hvh,gossiaux,rapp,surasree,santosh,dks}  
%but still it can be commented that the issue 
%is not settled yet. 
The values of the drag and diffusion coefficients can be used to characterize the distribution function of the probe.
Therefore, these transport coefficients can be utilized to understand the departure of the 
CQ distribution from the thermal distribution of the bath particles.
The shape of the equilibrium distribution function of CQ have been studied using a 
generalized Einstein relation derived in Ref.\cite{rafelski}. We revisit this relation by including 
both the collisional and the radiative transport coefficients of CQ.

The present work is organized as follows. In the next section, we discourse the formalism of Fokker 
Planck Equation(FPE) and the procedure to evaluate 
transport coefficients for collisional and radiative processes. 
In section III, the equilibrium distribution($f^{CQ}_{eq}$) of CQ is elaborated
in the context of bremsstrahlung process. The impact of the radiative transport coefficients on the 
$f^{CQ}_{eq}$ is particularly highlighted. In section IV, the $\eta/s$ of QGP has been estimated using CQ transverse diffusion
coefficients with particular emphasis on the radiative processes. Section V is dedicated to summary and conclusions.

\section{\bf Formalism and Transport Coefficients}
Heavy Quark propagates as a Brownian particle in the QGP medium. The ensemble of Brownian particles
immersed in the thermal medium can be characterized by the single particle distribution function, 
$f(\vec{x},\vec{p},t)$. The time evolution of $f$ is governed by the master equation, a simplified version of 
which is the Fokker Planck Equation(FPE). 

The form of the master equation or the Boltzmann transport equation(BTE) governing the CQ distribution $f$ is given by:
%%%%%%%%%%%%%%%%%%%%%%%%%%%%%%%%%%%%%%%%%%%%%%%%%%%%%%%%%%%%%%%%%%%%%%%%%
\bea
\left[\frac{\partial}{\partial t}+\frac{\vec{p}}{E}.\frac{\partial}{\partial \vec{x}}
+\vec{F}.\frac{\partial}{{\partial \vec{p}}}
\right] f(\vec{x},\vec{p},t)=\left[\frac{\partial f}{\partial t}\right]_{collisions}
\eea
%%%%%%%%%%%%%%%%%%%%%%%%%%%%%%%%%%%%%%%%%%%%%%%%%%%%%%%%%%%%%%%%%%%%%%%%%
In the absence of external force, $\vec{F}$ and for a homogeneous plasma, we can write BTE as follows:
%%%%%%%%%%%%%%%%%%%%%%%%%%%%%%%%%%%%%%%%%%%%%%%%%%%%%%%%%%%%%%%%%%%%%%%%%
\be
\left[\frac{\partial f}{\partial t}\right]_{collisions}= \int d^{3}\vec{k}[w(\vec{p}+\vec{k},\vec{k})
f(\vec{p}+\vec{k})-w(\vec{p},\vec{k})f(\vec{p})].
\label{BTE}
\ee
%%%%%%%%%%%%%%%%%%%%%%%%%%%%%%%%%%%%%%%%%%%%%%%%%%%%%%%%%%%%%%%%%%%%%%%%%%
where $w(\vec{p},\vec{k})$ is the rate of collision of CQ changing its momentum from $\vec{p}$ to $\vec{p}-\vec{k}$.
Considering only the soft scattering of CQ with the bath particles, we reduce the integro-differential Eq.~\ref{BTE}
to the FPE:
%%%%%%%%%%%%%%%%%%%%%%%%%%%%%%%%%%%%%%%%%%%%%%%%%%%%%%%%%%%%%%%%%%%%%%%%%
\be
\frac{\partial f}{\partial t}= \frac{\partial}{\partial p_{i}}\left[A_{i}(\vec{p})f+\frac{\partial}{\partial p_{j}}[B_{ij}
(\vec{p})f]\right]~~,
\label{landaukeq}
\ee
where the kernels are defined as
\be
A_{i}= \int d^{3}\vec{k}w(\vec{p},\vec{k})k_{i}~~,
\label{eqdrag}
\ee
and
\be
B_{ij}= \frac{1}{2} \int d^{3}\vec{k}w(\vec{p},\vec{k})k_{i}k_{j}.
\label{eqdiff}
\ee
%%%%%%%%%%%%%%%%%%%%%%%%%%%%%%%%%%%%%%%%%%%%%%%%%%%%%%%%%%%%%%%%%%%%%%%%%%
where $A_i$ and $B_{ij}$ are  the drag and the diffusion coefficients of CQ.
Our motivation is to find out these coefficients due to the elastic and inelastic 
interactions of the CQs with the bath particle within the ambit of pQCD.

\subsection{Transport co-efficient for collisional process}
First, we concentrate on the two-body elastic processes.  While propagating inside the plasma, the CQ(Q) 
encounters the following interactions with the bath particle: Q($p$)+g($q$)$\rightarrow $Q($p'$)+g($q'$),
Q($p$)+{\text q}($q$)$\rightarrow$ Q($p'$)+{\text q}($q'$) and
Q($p$)+$\bar{\text q}(q$)$\rightarrow$ Q($p'$)+$\bar{\text q}(q'$), where the quantities within the bracket denote momenta of
the CQ, quark ({\text q}), antiquark ($\bar{\text q}$) and gluon (g).   
%%%%%%%%%%%%%%%%%%%%%%%%%%%%%%%%%%%%%%%%%%%%%%%%%%%%%%%%%%%%%%%%%%%%%%%%%%%
%\be
%Q(p)+q/\bar{q}/g(q)\rightarrow Q(p')+q/\bar{q}/g(q')
%\ee
%%%%%%%%%%%%%%%%%%%%%%%%%%%%%%%%%%%%%%%%%%%%%%%%%%%%%%%%%%%%%%%%%%%%%%%%%%%
Therefore, $A_i$ and $B_{ij}$ are written in terms of invariant amplitude squared~\cite{BS} as:
%%%%%%%%%%%%%%%%%%%%%%%%%%%%%%%%%%%%%%%%%%%%%%%%%%%%%%%%%%%%%%%%%%%%%%%%%%%%%
\bea
A_i&=&\frac{1}{2E_{p}}\int \frac{d^3q}{(2\pi)^{3}2E_q}\int \frac{d^3q'}{(2\pi)^{3}2E_{q'}}\nn\\
&\times&\int \frac{d^3p'}{(2\pi)^{3}2E_{p'}}\frac{1}{\gamma}\sum |M|_{2\rightarrow2}^{2}(2\pi)^4\delta^4(p+q-p'-q')\nn\\
&\times&\hat{f}(\textbf{q})(1\pm\hat{f}(\textbf{q}'))(p-p')_i\\
&=&\ll(p-p')_i\gg
\label{ai}
\eea
%%%%%%%%%%%%%%%%%%%%%%%%%%%%%%%%%%%%%%%%%%%%%%%%%%%%%%%%%%%%%%%%%%%%%%%%%%%%%%%
Similarly,
%%%%%%%%%%%%%%%%%%%%%%%%%%%%%%%%%%%%%%%%%%%%%%%%%%%%%%%%%%%%%%%%%%%%%%5555
\be
B_{ij}=\frac{1}{2}\ll(p'-p)_{i}(p'-p)_{j}\gg
\label{bij}
\ee
%%%%%%%%%%%%%%%%%%%%%%%%%%%%%%%%%%%%%%%%%%%%%%%%%%%%%%%%%%%%%%%%%%%%%%%%%%%5
Since, $A_i$ and $B_{ij}$ only depend on three momentum, $\vec{p}$ and background temperature, T, we can write them as:
\be
A_i=p_i A
\label{Ai}
\ee
and
\be
B_{ij}=(\delta_{ij}-\frac{p_i p_j}{p^2})B_{\perp}(p,T)+\frac{p_i p_j}{p^2}B_{\parallel}(p,T).
\label{Bij}
\ee
where, $B_{\bot}$ and $B_{||}$ are transverse and longitudinal diffusion coefficients respectively.
Using Eqs.~\ref{ai},~\ref{bij},~\ref{Ai} and ~\ref{Bij} we get:
\bea
A(p)=\ll1\gg-\frac{\ll\vec{p}.\vec{p'}\gg}{p^2},
\label{a}
\eea
\bea
B_{\perp}(p)= \frac{1}{4}\left[\ll p'^2 \gg -\frac{\ll (\vec{p}.\vec{p'})^2 \gg}{p^2}\right],
\label{bperp}
\eea
\be
B_{\parallel}(p)= \frac{1}{2}\left[\frac{\ll (\vec{p}.\vec{p'})^2 \gg}{p^2}-2\ll \vec{p}.\vec{p'}\gg
+p^2 \ll1\gg\right].
\label{bpara} 
\ee
Using Eqs.~\ref{a},\ref{bperp} and \ref{bpara} with the matrix elements of those elastic processes mentioned above,
the drag, transverse and longitudinal diffusion coefficients can be estimated.

The expression for the transport coefficient($X(\vec{p},T))$ can be schematically written as:
\be
X(p)=\int phase~space\times interaction\times transport~part
\label{transport}
\ee
In case of drag(diffusion), transport part involves momentum(square of the momentum) transfer of CQ with the bath particle.
The evaluation of the drag and diffusion coefficients with collisional
loss are elaborated in Refs.~\cite{surasree,santosh}.
Therefore, we refer to these references for details and do not repeat the
discussions here. 
 
\subsection{Transport co-efficient for radiative process}
In view of Eq.~\ref{transport}, we can estimate radiative transport coefficients. There, the phase space
factor will be that for three-body scattering, the interaction will involve three-body invariant amplitude squared
and the transport part will remain same. The $X(\vec{p},T)$
for the radiative process, Q($p$)+parton($q$)$\rightarrow$  Q($p'$)+parton($q'$) +gluon ($k_5$), 
(where parton stands for light quarks, anti-quarks and gluons and $k_5=(E_5,k_{\perp},k_z)$) can be written as: 
\bea
X&=& \frac{1}{2E_p}\int \frac{d^3q}{(2\pi)^{3}2E_q}\int \frac{d^3q'}{(2\pi)^{3}2E_{q'}}
\int \frac{d^3p'}{(2\pi)^{3}2E_{p'}}
\nn\\&\times&\int \frac{d^3k_5}{(2\pi)^{3}2E_{5}}\frac{1}{\gamma}\sum |M|_{2\rightarrow3}^{2}
(2\pi)^4\delta^4(p+q-p'-q'-k_5)\nn\\
&\times&\hat{f}(E_q)(1\pm\hat{f}(E_{q'}))(1+\hat{f}(E_5))\nn\\
&\times& \theta(\tau-\tau_{F})\theta(E_p-E_5)
\label{radtransport}
\eea
where $\tau_F$
is the formation time of the emitted gluon, % (only the soft gluon emission will be considered here), 
step functions are introduced in Eq.~\ref{radtransport} to take into accounts
the LPM effects and to prohibit the emission of gluons with energy greater than $E_p$.
%In Eq.~\ref{radtransport}, LPM effect has been taken care of through $\theta_1$ and $\theta_2$ forbids the energy
%of the radiated gluon to be greater than that of the HQ. 
In order to calculate $\sum |M|_{2\rightarrow3}^2$, the necessary
Mandelstam variables are defined as:
\bea
s&=&(p+q)^2,~~~~s'=(p'+q')^2,\\
t&=&(p-p')^2,~~~~t'=(q-q')^2,\\
u&=&(p-q')^2,~~~~u'=(q-p')^2,
\label{mandelstam}
\eea
with
\be
s+t+u+s'+t'+u'=4M^2.
\ee
In this work, we will consider the case of soft gluon emission, i.e. when $k_5 \rightarrow 0$, 
which implies $s'\rightarrow s, 
t'\rightarrow t, u'\rightarrow u$.  For the kinematic region, 
\be
\sqrt{s}\gg \sqrt{|t|}\gg E_5
\ee
the invariant amplitude squared for $2\rightarrow 3$ process can be expressed
in terms of $2\rightarrow 2$ process multiplied with the emitted gluon spectrum as:
\be
|M|_{2\rightarrow3}^2=|M|_{2\rightarrow2}^2\times 12g_s^2 \frac{1}{k_{\perp}^2}\left(1+\frac{M^2}{s}e^{2y}\right)^{-2}
\label{radamplitude}
\ee
The last term in Eq.~\ref{radamplitude} is the dead cone factor~\cite{abir} and $y$ denotes the rapidity of the emitted gluon.
Following the prescription given in Eq.~\ref{radamplitude}, we have the radiative $X$:
\bea
X_{rad}&=&X_{coll}\times \int \frac{d^3k_5}{(2\pi)^{3}2E_5}12g_s^2\frac{1}{k_{\perp}^2}\nn\\
&\times&\left(1+\frac{M^2}{s}e^{2y}\right)^{-2}[1+\hat{f}(E_5)]\nn\\
&\times& \theta(\tau-\tau_{F})\theta(E_p-E_5)
\eea
After having calculated the radiative transport coefficients, we find out total or effective transport
coefficient as the sum of the collisional and radiative contribution, i.e.,
\be
X_{eff}=X_{coll}+X_{rad}.
\ee
where $X_{coll}$ and $X_{rad}$ are the transport coefficients for the collisional and radiative
processes respectively. 
In order to obtain the effective transport coefficient, we have added the collisional and radiative parts with
the view in mind that though the invariant amplitude of three body scattering can be expressed in 
terms of the two body scattering, the process of collision and radiation are taking place inside the
thermal medium independently. 

%All the transport coefficients are functions of momentum
%of CQ and temperature of the heat bath. Therefore, effective and collisional transport 
%coefficients are plotted once with temperature keeping momentum of charm quark as a parameter and 
%then with momentum of charm holding temperature as parameter.

%%%%%%%%%%%%%%%%%%%%%%%%%%%%%%%%%%%%%%%%%%%%%%%%%%%%%%%%%%%%%%%%%%%%%5
\begin{figure}[h]
\begin{center}
\includegraphics[scale=0.43]{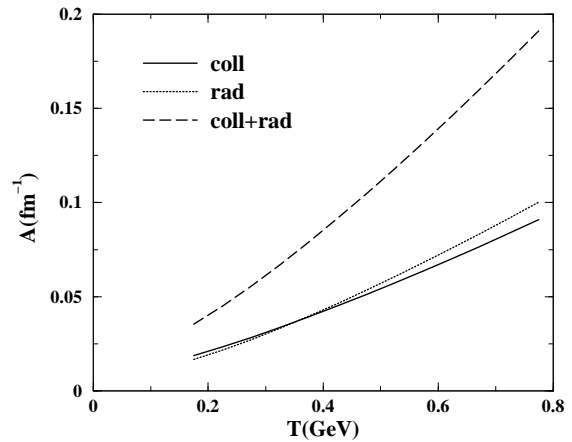}
\caption{Temperature dependence of the drag coefficient of CQ of momentum, $p=5$  GeV.}
\label{fig1}
\end{center}
\end{figure}
%%%%%%%%%%%%%%%%%%%%%%%%%%%%%%%%%%%%%%%%%%%%%%%%%%%%%%%%%%%%%%%%%%%%%%%
\begin{figure}[h]
\begin{center}
\includegraphics[scale=0.43]{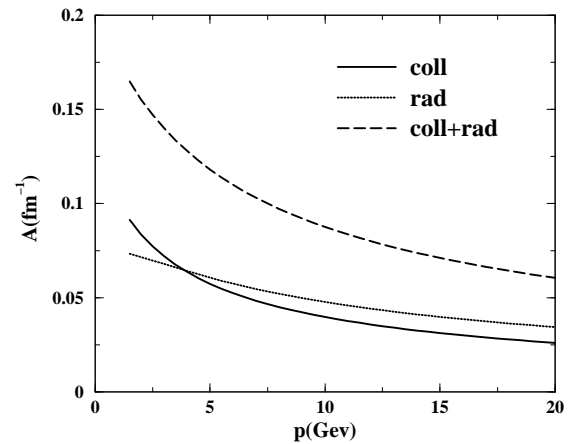}
\caption{Momentum dependence of the drag coefficient of CQ for bath temperature, $T=525$ MeV}
\label{fig2}
\end{center}
\end{figure}
%%%%%%%%%%%%%%%%%%%%%%%%%%%%%%%%%%%%%%%%%%%%%%%%%%%%%%%%%%%%%%%%%%%%%%%%
\begin{figure}[h]
\begin{center}
\includegraphics[scale=0.43]{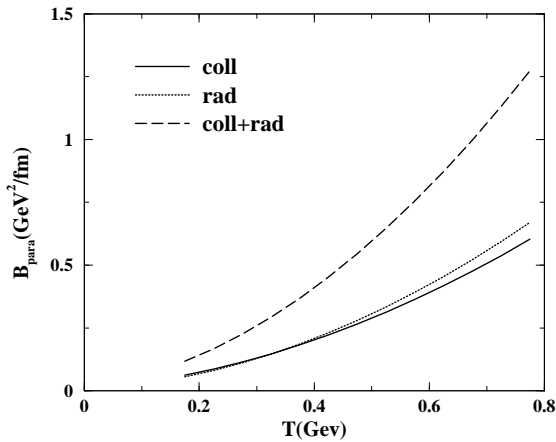}
\caption{Temperature dependence of the longitudinal diffusion coefficient of CQ of momentum, $p=5$  GeV.}
\label{fig3}
\end{center}
\end{figure}
%%%%%%%%%%%%%%%%%%%%%%%%%%%%%%%%%%%%%%%%%%%%%%%%%%%%%%%%%%%%%%%%%%%%%%%%%%5
\begin{figure}[h]
\begin{center}
\includegraphics[scale=0.43]{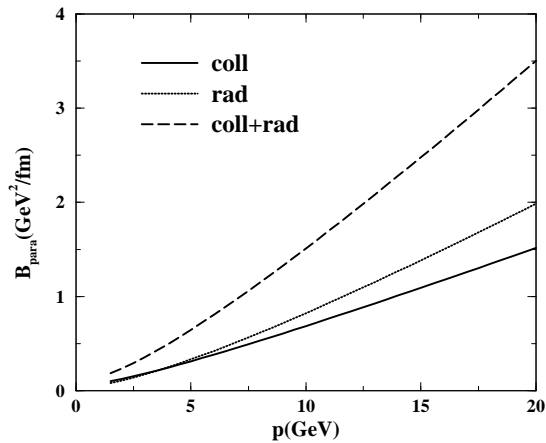}
\caption{Momentum dependence of the longitudinal diffusion coefficient of CQ for bath temperature, $T=525$ MeV}
\label{fig4}
\end{center}
\end{figure}
%%%%%%%%%%%%%%%%%%%%%%%%%%%%%%%%%%%%%%%%%%%%%%%%%%%%%%%%%%%%%%%%%%%%%%%%
\begin{figure}[h]
\begin{center}
\includegraphics[scale=0.43]{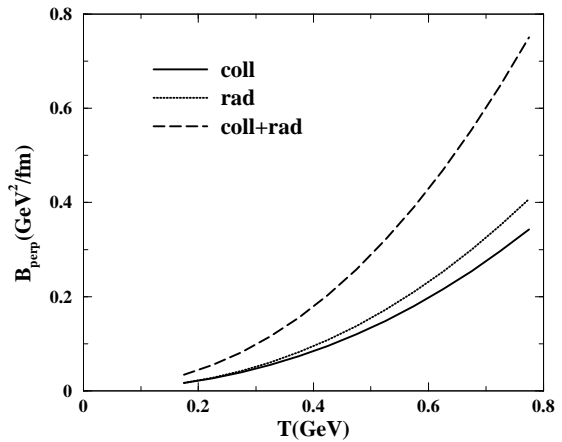}
\caption{Temperature dependence of the transverse diffusion coefficient of CQ of momentum, $p=5$  GeV.}
\label{fig5}
\end{center}
\end{figure}
%%%%%%%%%%%%%%%%%%%%%%%%%%%%%%%%%%%%%%%%%%%%%%%%%%%%%%%%%%%%%%%%%%%%%%%%%
\begin{figure}[h]
\begin{center}
\includegraphics[scale=0.43]{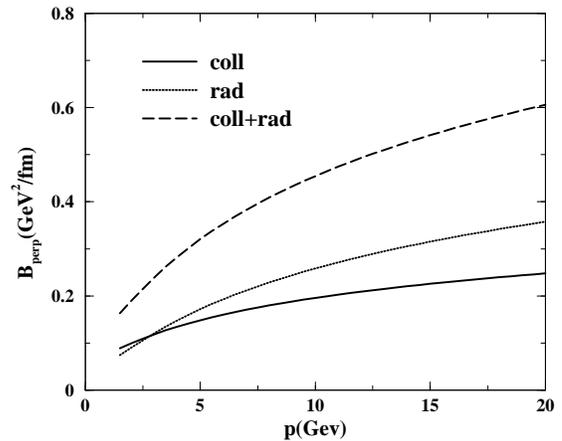}
\caption{Momentum dependence of the transverse diffusion coefficient of CQ for bath temperature, $T=525$ MeV}
\label{fig6}
\end{center}
\end{figure}
%%%%%%%%%%%%%%%%%%%%%%%%%%%%%%%%%%%%%%%%%%%%%%%%%%%%%%%%%%%%%%%%%%%%%%%%%%

In Fig.~\ref{fig1} we display the temperature dependence of the drag of CQ 
with momentum, $p=5$ GeV. At low $T$, although, the drag for radiative loss
is comparable to the collisional loss, at high temperature the radiative
drag tend to dominate. 
%The temperature dependence of all the three transport coefficients are similar, i.e. they increase with
%temperature. One can very well observe from the plots that the radiative transport coefficients are also
%of same order as that of the collisional. Therefore, total or effective coefficients are well above in magnitude
%than that of the collisional only. 
The difference between total and collisional transport coefficients
broadens with increasing temperature. Even at temperatures attainable in RHIC, this distinction is 
significant enough to have a pronounced effect on certain experimental observables like nuclear modification 
factor, elliptic flow of CQ etc.  In the temperature range that may be achieved at LHC collision
conditions the radiative contributions to the drag may surpass the collisional contributions.
%At LHC energies, the temperature of the bath may be such that the radiative transport 
%coefficients surpass collisional coefficients in magnitude. 
Therefore, radiative processes will
play more dominant role at LHC than at RHIC. For a           
CQ (mass, $M=1.3$ GeV) with $p=5$ GeV and $T=300$ MeV, the value of drag coefficient attains a value almost double the value for 
collisional case when radiation is included. At 600 MeV temperature, total drag becomes 2.12
times the collisional drag. 
The variation of drag with $p$ at $T=525$ MeV is depicted in Fig.~\ref{fig2}. The dominance of radiative processes,
in spite of dead cone suppression, is evident from the results  for $p$ beyond 5 GeV.

In Fig.~\ref{fig3} and \ref{fig4}, the variation of longitudinal diffusion coefficient with
temperature and momentum are displayed respectively. Similar to drag, the contributions from radiative processes
dominate over the collisional processes for higher $T$ and $p$.   

For $T=300$ MeV, the radiative and collisional loss have similar contributions to $B_{\perp}$, 
but for $T=600$ MeV the radiative part exceeds the collisional part (Fig~\ref{fig5}). 
It is interesting to note the qualitative change in the momentum dependence of 
$B_{\perp}$   from $B_{\parallel}$ at fixed $T$ (Fig.~\ref{fig6}). The variation of
$B_{\perp}$ with $p$ is slower than $B_{\parallel}$. 
In this case again the domination of the radiative transport coefficient over its collisional counterpart is evident.
Though the nature of the momentum dependence of the 
diffusion coefficients is different  from that of drag, it is always true that, save at very low momentum of
the CQ, radiative contribution is more than the elastic contribution at $T=525$ MeV 
for the momentum range considered here. 
Accordingly, for a relativistic CQ the effect of radiation becomes imperative and
should be included in the analysis of experimental data from nuclear collisions 
at RHIC and LHC. This statement can be kept on a firmer ground if we quote some quantitative 
results comparing radiative and collisional contributions to the transport coefficients. 
Drag coefficient of a CQ having momentum 10 GeV
is 0.038$fm^{-1}$ in case of elastic loss, whereas the radiative contribution is 0.047$fm^{-1}$. 
Radiative $B_{\perp}$ is about 1.33 times its collisional counterpart. In case of longitudinal
diffusion coefficient, radiative contribution is 1.2 times the elastic one.

\section{\bf Equilibrium distribution of charm quark}
Having calculated the diffusion coefficients of CQs including radiative effects, we would like to
investigate  the fate of 
the equilibrium distribution function of a 
CQ undergoing elastic as well as radiative processes. A generalized Einstein relation
involving the three transport coefficients, {\it i.e.} drag, transverse and longitudinal diffusion coefficients  
is obtained in \cite{rafelski} to establish the shape of the distribution of CQ after it gets equilibrated due to its collisional and
radiative interaction with the medium. In Ref.~\cite{rafelski} the radiative process was not taken into account.
We would like to explore the role of the radiative processes in the characterisation
of the equilibrium distribution and to check whether CQ becomes a part of the thermal medium abiding by the same class 
of statistics, which is the Boltzmann-J\"{u}ttner distribution, followed by the bath particles.

We discuss the generalized Einstein relation
by examining the Fokker Planck equation in the absence of any external force in a homogeneous QGP(Eq.~\ref{probability}).
\bea
\frac{\partial f}{\partial t}=\frac{\partial}{\partial p_{i}}\left(A_if+\frac{\partial}{\partial p
_{j}B_{ij}f}\right)=-\vec{\nabla_{p}}.\vec{\wp}
\label{probability}
\eea  
A relationship among the transport coefficients can be derived by demanding that $\partial f/\partial t$ is zero,
i.e. the probability current, $\vec{\wp}$ vanishes when Eq.~\ref{probability} is satisfied by the equilibrium distribution
function, $f^{CQ}_{eq}$.

Using the following form of $f^{CQ}_{eq}$, 
\be
f^{CQ}_{eq}(p;T,q)=Nexp[-\Phi(p;T,q)]
\label{equidistribution}
\ee
the desired relation can be found out,
where, N is the normalization factor; $T$, $q$ are parameters needed to specify
the shape of the distribution. Using Eqs.~\ref{Ai} and \ref{Bij} and the fact that $f^{CQ}_{eq}$ depends 
only on the magnitude of momentum for spatially homogeneous case, we arrive at the general Einstein relation:
\bea
A(p,T)=\frac{1}{p}\frac{d\Phi}{dp}B_{\parallel}(p,T)-\frac{1}{p}\frac{dB_{\parallel}}{dp}\nn\\-
\frac{2}{p^2}[B_{\parallel}(p,T)-B_{\perp}(p,T)]
\label{einsteinreln}
\eea 
This relation is valid for any momentum of CQ and can be reduced to the well-known 
Einstein relation, $D=\gamma M T$, 
in the non-relativistic limit, where  $A=\gamma$ and 
$B_{\perp}=B_{\parallel}=D$, i.e. $B_{ij}=D\delta_{ij}$ and $\Phi=p^2/(2MT)$. 

\begin{figure}[h]
\begin{center}
\includegraphics[scale=0.43]{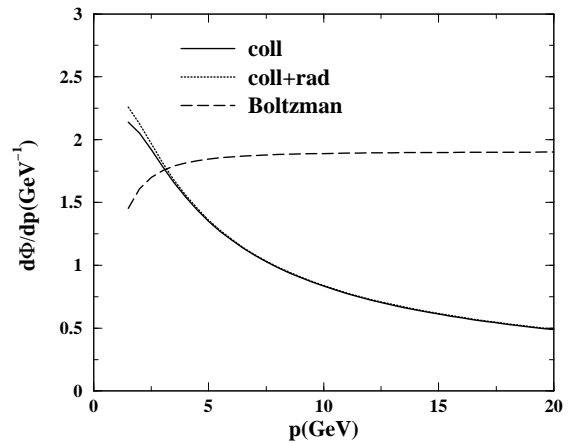}
\caption{For CQ propagating in QGP having temperature, $T=525$ MeV.}
\label{fig7}
\end{center}
\end{figure}

\begin{figure}[h]
\begin{center}
\includegraphics[scale=0.43]{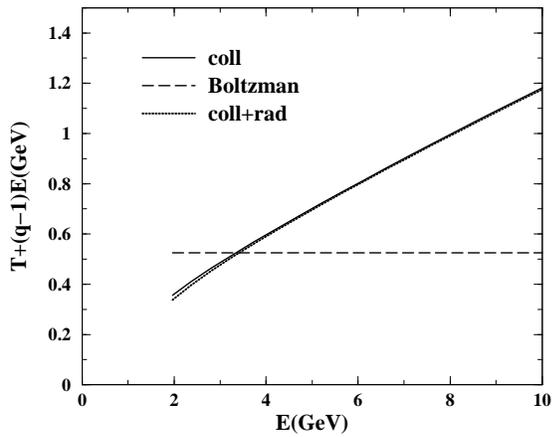}
\caption{Plot of RHS of Eq.~\ref{Tq} vs $E$ for collisional as well as total transport coefficients at
$T=525$ MeV. Long dashed line: expected for Boltzmann-J\"{u}ttner distribution}
\label{fig8}
\end{center}
\end{figure}

From Eq.~\ref{einsteinreln}, it is quite conspicuous that if the three drag/diffusion coefficients are known
then one can infer the correct equilibrium distribution function obeyed by CQ and ascertain whether or not
CQ will fall under Boltzmann-J\"utner class of statistics. 
It is clear from the variation of  $d\Phi/dp$  (calculated from Eq.~\ref{einsteinreln})  with  $p$
(Fig.~\ref{fig7}) that $d\Phi/dp$ deviates significantly from $d/dp(\sqrt{p^2+m^2}/T)$,
%%%%%%%%%%%%%%%%%%%%%%%%%%%%%%%%%%%%%%%%%%%%%%%%%%%%%%%%%%%%%%%%%%%%%%%%%%%%%%%%%%%%
%%%%%%%%%%%%%%%%%%%%%%%%%%%%%%%%%%%%%%%%%%%%%%%%%%%%%%%%%%%%%%%%%%%%%%%%%%%%%%%%%%55
i.e. CQ seems to be away from the Boltzmann-like distribution.
In principle, we should have ascertained the precise form of $\Phi$ from Eq.~\ref{einsteinreln},
had we been able to include all the non-perturbative contributions to calculate $A$, $B_{\perp}$ and $B_{\parallel}$.
To study the equilibrium distribution 
and its deviation from the Boltzmann-J\"{u}ttner distribution quantitatively, we consider 
Tsallis distribution~\cite{tsallis} for which $\Phi$ is given by:
\be
\Phi_{Ts}=\frac{1}{1-q}ln\left[1-(1-q)E(p)/T_T\right]~,
\label{phits}
\ee  
where $T_T$ (temperature like) and  q are parameters. The $\Phi_{Ts}$ reduces to the Boltzmann distribution 
in the limit $q\rightarrow 1$ and $T_T\rightarrow T$($T$ is the temperature 
of the heat bath). The value of $T_T$ and $q$ will designate the 
form of $f^{CQ}_{eq}$. Putting Eq.~\ref{phits} into Eq.~\ref{einsteinreln}, we get\cite{rafelski} 
\be
T_T+(q-1)E=\frac{dE}{dp}\frac{B_{\parallel}}{pA+\frac{dB_{\parallel}}{dp}+\frac{2}{p}(B_{\parallel}-B_{\perp})}
\label{Tq}
\ee
Our aim is to calculate the right hand side of Eq.~\ref{Tq} and to determine  the value of $T_T$ and $q$
by studying the variation of  $T_T+(q-1)E$ with  $E$  and parameterizing the variation by 
straight line.  First we consider the elastic processes only.
The dependence of  $T_T+(q-1)E$ on $E$ for CQ of mass, $M=1.3$ GeV propagating inside a heat bath
having temperature, $T=525$ MeV is plotted in Fig.~\ref{fig8} considering $A$, $B_{\perp}$ and $B_{\parallel}$ 
for collisional loss only. We get $q=1.101$ and $T_T=184$ MeV. $\Phi_{Ts}$
with these value of $T_T$ and $q$ is far from being that of the of Boltzmann-J\"{u}ttner statistics (long-dashed line).
%%%%%%%%%%%%%%%%%%%%%%%%%%%%%%%%%%%%%%%%%%%%%%%%%%%%%%%%%%%%%%%%%%%%%%%%%%%%%%%%
%%%%%%%%%%%%%%%%%%%%%%%%%%%%%%%%%%%%%%%%%%%%%%%%%%%%%%%%%%%%%%%%%%%%%%%%%%%%5555  
Results displayed in Fig.~\ref{fig8} also indicates that the inclusion of radiative effects on the drag and diffusion coefficients does not 
make any noteworthy change on the shape of the equilibrium distribution of CQ. With the total transport coefficients,
the values of $T_T$ and q do not get altered. As a matter of fact,
this effect is not quite unexpected.
By looking at  Eq.~\ref{Tq}, we might conclude that it is not the magnitude of the transport coefficients, rather their
ratio which decides the shape of the equilibrium distribution. Therefore, it is not surprising at all that 
although the value of the relaxation time of CQ is dictated by the magnitude of the drag 
coefficient (in which the radiative contribution is substantial), 
the shape is largely independent of the magnitude of the transport coefficients. In turn, this means that the nature
of the underlying interaction of CQ with the bath particles, i.e. whether it suffers only elastic collisions
or it undergoes bremsstrahlung also, has got to do very little with the ultimate shape of $f^{CQ}_{eq}$. 
%Consequently, the shape is, so to say, history independent in the above context.
This conclusion remains  unaltered even when we increase the bath 
temperature, $T$ to 725 MeV.
%%%%%%%%%%%%%%%%%%%%%%%%%%%%%%%%%%%%%%%%%%%%%%%%%%%%%%%%%%%%%%%%%%%%%%%%%%%%%%%%%%
\begin{figure}[h]
\begin{center}
\includegraphics[scale=0.43]{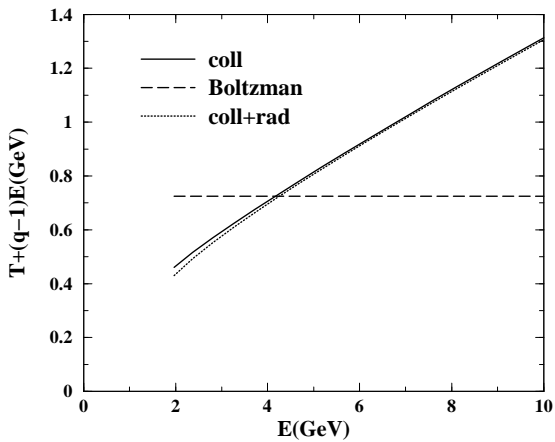}
\caption{Plot of RHS of Eq.~\ref{Tq} vs $E$ for collisional as well as total transport coefficients at
$T=725$ MeV. Long dashed line: expected for Boltzmann-J\"{u}ttner distribution}
\label{fig9}
\end{center}
\end{figure}
%%%%%%%%%%%%%%%%%%%%%%%%%%%%%%%%%%%%%%%%%%%%%%%%%%%%%%%%%%%%%%%%%%%%%%%%%%%%%%%%%5
At $T=725$ MeV, the value of the slope of the straight line remains unchanged, i.e. 
the value of the parameter $q$ comes out to be 1.095 which is almost same as that at $T=525$ MeV. But, the value 
of the other parameter of the Tsallis distribution, $T_T$ is found out to be 335 MeV. At this temperature
of the heat bath too, the incorporation of the radiative drag/diffusion coefficients hardly has any 
bearing as far as the shape of $f^{CQ}_{eq}$ is concerned. 
In Fig.~\ref{fig8} and Fig.~\ref{fig9}, the long-dashed horizontal lines represent the Boltzmann-J\"{u}ttner
distribution($q=1$ and $T_T=T$) which is obeyed by the constituent of QGP. In order to expect the probe
particle, the CQ, to become a part of the system, i.e. to follow the same statistics as that of the bath 
particles, one should have found out the parameters q and $T_T$ of the Tsallis distribution to be 1 and $T$ respectively.
Instead, we notice that q and the ratio, $T/T_T$ (this ratio is 2.85 at $T=525$ MeV and 2.164 at $T=725$ MeV) are
never equal to unity. 
Therefore, it may be concluded that though, the CQ may equilibrate while propagating through QGP, it may not share
the same distribution with the bath particles, i.e. with the light quarks and gluons for a wide 
range of CQ energies and bath temperatures.

\section{Shear viscosity($\eta$) to entropy density($s$) ratio of QGP probed by charm quark}
The value of shear viscosity($\eta$) to entropy density($s$) ratio, $\eta/s$, play a pivotal role in
deciding the nature of QGP, i.e. whether the medium behaves like a  weakly coupled gas or a strongly coupled liquid. 
%The magnitude of
%$\eta/s$ has been estimated rlier via $\pi^0$ suppression measurements and flow measurements\cite{laceyprl98}.
In this work  we evaluate $\eta/s$ by calculating the transport parameter, $\hat{q}$, which is a measure of the squared
average  momentum exchange between the probe and the bath particles per unit length~\cite{amajumderprl99,baiernpb,laceyprl103}.
 %The $\hat{q}$
%has been found to be $\sim$ 1 $GeV^2/fm$\cite{laceyprl103}. 
The $\hat{q}$, which has been found to be $\sim$ 1 GeV$^2$/fm in Ref.~\cite{laceyprl103}, can be related to
the transverse diffusion coefficient of CQ which is calculated here. 
When a CQ with a certain momentum propagates in QGP, a 
transverse momentum exchange with the bath particles occurs. Hence, the 
momentum of the energetic CQ is shared by the low momentum (on the average) bath particles which
is expressed through the transverse diffusion coefficients. The transverse 
diffusion coefficients causes the minimization of the momentum  (or velocity) gradient  
in the system. Therefore, it must be related to the shear viscous coefficients 
of the system which drive the system towards depletion of the velocity gradient.
The transverse momentum diffusion coefficient, $B_{\perp}$ can be written as:
\be
B_{\perp}=\frac{1}{2}\left(\delta_{ij}-\frac{p_ip_j}{p^2}\right)B_{ij}
\ee
By Eq.~\ref{bij} and using the notation $(p'-p)_i=k_i$, 
\bea
B_{\perp}&=&\frac{1}{2}\left(\delta_{ij}-\frac{p_ip_j}{p^2}\right)\frac{1}{2}\ll k_ik_j \gg \nn\\
&=&\frac{1}{4}\ll \vec{k}^2-\frac{(\vec{p}.\vec{k})^2}{\vec{p}^2}\gg \nn
\eea
If we take $\vec{p}$ to be along z-axis,
\bea
B_{\perp}&=&\frac{1}{4}\ll \vec{k}^2-k_z^2\gg\nn\\
&=&\frac{1}{4}\ll k_{\perp}^2\gg\nn\\
&=&\frac{1}{4}\hat{q} 
\eea
With this definition of $\hat{q}$, we calculate $\eta/s$ of QGP from 
the following expression\cite{amajumderprl99}:
\be
\frac{\eta}{s}\approx 1.25\frac{T^3}{\hat{q}}
\ee
Therefore,
\be
4\pi\frac{\eta}{s}\approx 1.25\pi\frac{T^3}{B_{\perp}}.
\label{etabys}
\ee
Eq.~\ref{etabys} indicates that the $\eta/s$ can be estimated from $B_{\perp}$. 
From the analysis of the experimental data~\cite{laceyprl103} it was found 
that $4\pi\frac{\eta}{s}=1.4\pm 0.4$
which may be compared with the AdS/CFT bound $4\pi\frac{\eta}{s}\geq 1$\cite{kovtumprl}. We display 
$4\pi\frac{\eta}{s}$ against $T$  when CQ undergoes both collisional 
and radiative processes. 
%%%%%%%%%%%%%%%%%%%%%%%%%%%%%%%%%%%%%%%%%%%%%%%%%%%%%%%%%%%%%%%%%%%%%%%%%%%%%%%%%%
\begin{figure}[h]
\begin{center}
\includegraphics[scale=0.43]{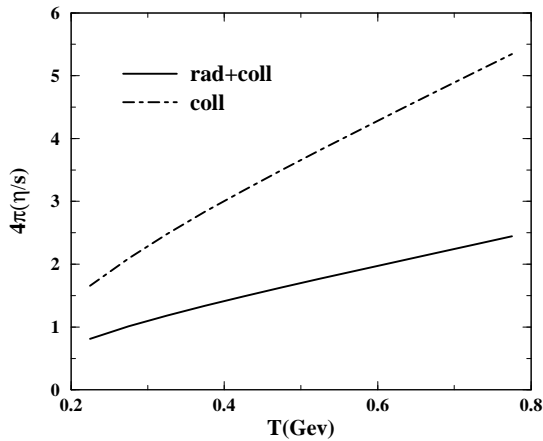}
\caption{For a CQ with momentum, $<p_T>=5$ Gev propagating in QGP of temperature, T}
\label{fig10}
\end{center}
\end{figure}
%%%%%%%%%%%%%%%%%%%%%%%%%%%%%%%%%%%%%%%%%%%%%%%%%%%%%%%%%%%%%%%%%%%%%%%%%%%%%%%%%% 

From the results shown in Fig.~\ref{fig10}, it should be noted that the value of $\eta/s$
changes substantially with the inclusion of the radiative effects. The inclusion 
of the radiative loss in $B_{\perp}$ brings the theoretical values closer to the experimental findings\cite{adareprl}. 
This highlights the importance of the radiative loss of the CQ in QGP. 
%A closer look at the results
%shown in Fig.~\ref{figetabys} by solid line which  corresponds to the values of $4\pi\eta/s$  
%obtained by including both the processes indicate that there are values below the quantum limit at low temperature
%region. This might be sensitive to the $<p_T>$ (which is taken to be 5 GeV here) of charm distribution and also
%to the accuracy of the pQCD expansions from which $B_{\perp}$ has been computed. 
\section{\bf Summary and conclusion}
Transport coefficients {\it i.e.}  drag, transverse and longitudinal diffusion coefficients of CQ propagating
in QGP have been evaluated using  pQCD by including both the elastic collision of CQ with the constituent
particles of the bath along with soft gluon radiation 
(which gets absorbed in the medium subsequently).
Radiative drag/diffusion coefficients are seen to exceed the collisional ones for high bath temperature and large
CQ momentum.
A relation between the transverse diffusion coefficients($B_{\perp}$) and $\eta/s$ is established. We obtain reasonable
value of $\eta/s$ for the  QGP when  the contributions from the gluon bremsstrahlung of the CQ
is added with the collisional contributions.
We also investigate the dependence of the shape of the equilibrium distribution function of CQ on the three
transport coefficients. We find that the incorporation of radiation does not alter the shape of the
equilibrium distribution significantly, 
 owing to the fact that the shape counts on the ratios of the transport coefficients instead of their 
absolute values.
The present work has been performed for CQ. However, its extension for bottom quark is straight forward,
where the mass of the charm quark has to be replaced by that of the bottom quark.

\section*{\bf ACKNOWLEDGMENT}
S.M. and T.B. are grateful to R. Abir for useful discussions and to the 
Department of Atomic Energy, Government
of India for financial support.

\end{document}